\documentclass[aps,prl,twocolumn,showpacs]{revtex4}
\usepackage{graphicx}
\def\maj#1{\ifmmode\mbox{\usefont{U}{msb}{m}{n}#1}\else{\usefont{U}{ms
b}{m}{n}#1}\fi}
\def\v#1{\mathbf{#1}}
\usepackage{graphics,epsfig,graphicx}
\usepackage{color}
\usepackage{latexsym}

\begin{document}

\draft \title{Bose-Einstein condensation in semiconductors: the key role of dark excitons}
\author{Monique Combescot$^{a}$, Odile Betbeder-Matibet$^{a}$ and Roland Combescot$^{b}$}
\address{(a) Institut des NanoSciences de Paris,Universit\'e Pierre et Marie Curie, CNRS,
Campus Boucicaut, 
140 rue de Lourmel, 75015 Paris}
\address{(b) Laboratoire de Physique Statistique, Ecole Normale
Sup\'{e}rieure, 24 rue Lhomond, 75005 Paris}
\date{Received \today}

\begin{abstract}
The non elementary-boson nature of excitons controls Bose-Einstein condensation in semiconductors. Composite excitons interact predominantly through Pauli exclusion; this produces dramatic couplings
between bright and dark states. In microcavities, where bright excitons and photons form polaritons, they force the condensate to be linearly polarized--as observed. In bulk, they also force linear polarization, but of dark states, due to interband Coulomb scatterings. To evidence this dark condensate, we thus need indirect processes, like the shift it induces on the (bright) exciton line.
\end{abstract}
\maketitle

Although the possibility for a system of non interacting massive bosons to condense into
a single coherent state was predicted almost one hundred years ago
by A. Einstein \cite{1}, it has taken quite a long time for such a
spectacular macroscopic quantum effect to be directly seen in weakly interacting
systems. Bose-Einstein condensation
(BEC) requires the quantum regime to be reached, with a de Broglie wavelength
comparable to the interparticle distance. This implies a very low temperature,
the heavier the bosons, the lower the temperature. The development in the 80's of elaborate cooling techniques in atomic physics has allowed the observation of
this BEC first in atomic gases \cite{2} and more recently in
molecular systems \cite{4} at temperatures as low as $\mu $K. These
observations have generated a tremendous interest and opened a new field
\cite{2} of very active research on these ``ultracold gases''.

Since it is favorable to have light particles to observe BEC, 
semiconductors are very appealing. Excitons,
which result from the Coulomb attraction of one conduction electron and
one valence hole, are known to exhibit a bosonic behavior. They should thus display Bose-Einstein condensation.
As their effective mass is of the order of the vacuum electron mass, they
are much lighter than atoms, so they should
condense at much higher temperature. This is why exciton BEC has been
searched for tens of years,
with repeated claims of observation followed by denials \cite{6}. Note that, being semiconductor excitations, excitons of course have a finite lifetime. However, their
lifetime seems to be long enough to reach the thermodynamical
equilibrium required for BEC.

Quite recently, a conceptually similar effect
has been convincingly observed through the Bose-Einstein condensation of
polaritons \cite{7,8}, which are excitations resulting from the strong coupling of one photon and
one exciton. In a microcavity, the extremely steep dispersion of the
polariton modes results in an extremely light polariton effective mass,
of the order of $10^{-4}$ times the vacuum electron mass. This allows
critical temperatures \cite{9} for polariton BEC as high as a few hundred K. While 2D
systems are known to never develop true long range order, coherence across
a finite size polariton cloud in a trapped geometry, with a macroscopic occupation of a single
quantum state, has been demonstrated \cite{8}.

In order to detect the exciton BEC, attention has been naturally focused on bright
excitons which are the ones coupled to light. This has led to overlook dark excitons which
do not have this advantage. In this letter, we point out the key role they play for BEC
in semiconductors. 

Selection among the possible Bose-Einstein condensates results from interactions.
The dominant one for
excitons is the Pauli exclusion between their fermionic components. It induces dramatic
exchange couplings between bright and dark states which forces the polariton condensate
to have a linear polarization, as it produces the lowest energy,  in full agreement
with experiments \cite{7,8}. In the case of exciton BEC, these bright-dark couplings also force the 
linear polarization of the condensate. However, the dark excitons having an energy slightly
lower than the bright excitons, due to valence-conduction Coulomb processes, the stable
condensate is actually made of dark excitons. This may well explain why the exciton BEC has not
yet been observed. This letter should thus stimulate conceptually new experiments looking for a dark
BEC. We suggest an indirect way to observe it through the shift of the
exciton line, induced by the interactions of a bright exciton with the dark excitons which have
condensed.

When the coupling between photons and excitons 
is weak compared to the exciton broadening, excitons are created by photon absorption 
according to the Fermi golden
rule \cite{10}. When it is strong, photons are not absorbed, but form a mixed state with the bright excitons. The resulting
exciton-polaritons \cite{11} are exact eigenstates of the coupled
photon-semiconductor Hamiltonian for one excitation. Since photons do not interact,
many-body effects between polaritons can only come from interactions between
the excitonic parts of these polaritons.

As transparent from the exciton many-body theory we have
recently constructed \cite{12,13,13b,14}, the interactions between excitons are
predominantly due to Pauli exclusion between their fermionic components,
departing in this way from the standard elementary boson picture. As shown in various
works \cite{15,16,17,18}, all optical nonlinear effects in semiconductors are 
driven by Pauli exclusion, Coulomb interactions only producing
corrective terms. 
The associated carrier exchanges, which
can take place between many more than two excitons, are nicely
visualized through Shiva diagrams \cite{19}. These diagrams not
only help to select the dominant processes in the physical effect at
hand, but also allow to calculate it readily. The key points of this
many-body theory can be found in \cite{13}. 

The actual Bose-Einstein condensate in semiconductors is selected in the degenerate exciton subspace
through many-body effects between excitons, since the condensate with the lowest energy is the one 
which is realized. Hence, the composite nature of the excitons through Pauli 
exclusion between its constituents plays a key role in this selection. 

\begin{figure}
\vspace{25mm}
%\vspace{-5mm}
%\scalebox{0.8}{\includegraphics[width=10cm]{Fig1bis}}
%\hspace{10mm}
\scalebox{1.}{\hspace{10mm} \includegraphics[width=10cm]{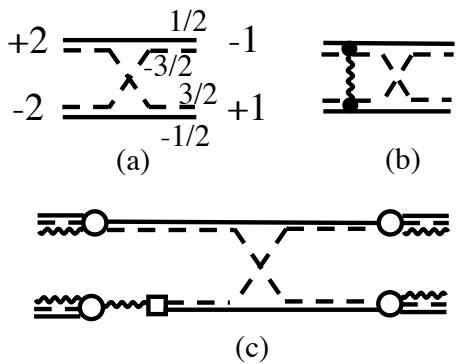}}
\vspace{-45mm}
\caption{(a) Pauli scattering for carrier exchange between two bright 'in'
excitons $(\mp 1)$. The 'out' excitons with spins $(\pm 2)$ are dark. Solid line: electron;
Dashed line: hole; Double solid-dashed line: exciton
(b) Exchange Coulomb scattering between two excitons. (c)~Photon assisted exchange scattering between two polaritons (triple solid-dashed-wavy line); Wavy line: photon; Empty square: exciton-photon coupling; Empty dot: couplings between polariton and exciton or photon}
%\vspace{-6mm}
\end{figure}
In quantum
wells, the exciton degeneracy is 4, the conduction electron having a
spin $s=\pm 1/2$ and the valence-hole an angular momentum
$m=\pm 3/2$, the light holes, with angular momentum $m=\pm 1/2$, having a higher
energy in confined geometry. Among these four exciton states, the two with total spin $\pm 2$ are not coupled to light; so they are not concerned with the polariton BEC. Consequently, to
characterize the polariton condensate, we must
find the linear combination of the two bright excitons leading to the
lowest energy, i.e. the prefactors $a_{\pm 1}$ of
\begin{equation}
B^\dag=a_1B_1^\dag+a_{-1}B_{-1}^\dag\ ,
\end{equation}
where $B_{\pm 1}^\dag$, defined as
\begin{equation}
B_{\pm 1}^\dag=\sum_{\v k}\langle\nu_0|\v k\rangle\,a_{\v k,\mp 1/2}^\dag
b_{-\v k,\pm 3/2}^\dag\ ,
\end{equation}
creates a $S=\pm 1$ ground state bright exciton with zero center-of-mass
momentum, in the relative motion ground state $|\nu_0\rangle$ with energy $-E_0$, made of a
$(\v k,s=\mp 1/2)$ electron and a $(-\v k,m=\pm 3/2)$ hole.

As obvious from Fig.1(a), the Pauli scattering for carrier exchanges
of two bright excitons $(\pm 1)$ generates two dark
excitons $(\pm 2)$. This escape into dark states reduces all
diagonal matrix elements between $n$ bright excitons $B^\dag$ by a factor
$(|a_1|^{2n}+|a_{-1}|^{2n})$, as readily seen \cite{21} from the Shiva
diagrams of Fig.2. Since
$|a_1|^2+|a_{-1}|^2=1$ for normalization, so that
$a_1=e^{i\varphi_1}
\cos\xi_1$ and $a_{-1}=e^{i\varphi_{-1}}\sin\xi_1$, where $\xi_1$ is
the ellipticity of the exciton linear combination $B^\dag$ , this
factor reduces  \cite{other} for $n=2$ to
\begin{eqnarray}\label{an}
|a_1|^{4}+|a_{-1}|^{4}&=&1 - 2|a_1a_{-1}|^{2}= 1-\frac{1}{2}\sin^22\xi_1
\end{eqnarray}

According to \cite{21}, this leads to an energy expectation value for $N$ excitons given by
\begin{eqnarray}\label{hn}
\hspace{-2mm} \langle H\rangle_N &=& \frac{\langle v|B^N\,H\,B^{\dag N}|v\rangle}
{\langle v|B^N\,B^{\dag N}|v\rangle}\nonumber\\
&=&
NE_0\left[-1+\eta\left(1-\frac{1}{2}\sin^22\xi_1\right)x_1+O(\eta^2)
\right]\;\;
\end{eqnarray}
where $\eta=N(a_\mathrm{x}/L)^2$ is the dimensionless parameter associated to the exciton
density, $a_x$ being the exciton Bohr radius, $L$ the sample size and
$x_1 = \pi-315\pi^3/4096\simeq 0.75$ in 2D \cite{22}. The $\eta$ term in Eq.(\ref{hn}) comes from the exchange Coulomb scattering between $n=2$ excitons
shown in Fig.1(b). It thus appears with the reduction factor for carrier
exchanges given in Eq.(\ref{an}). Note that the
direct Coulomb scattering reduces to zero when the repulsion between identical
fermions is as strong as the attraction between different fermions --- as
for Coulomb interaction (see Eq. (B18) of \cite{13b}). Eq.(\ref{hn}) thus shows that the minimum energy is reached for $\xi_1=\pi/4$,
\emph{i.e.}, for \emph{linearly polarized} excitons.
\begin{figure}
\vspace{-35mm}
%\hspace{-7mm}
%\scalebox{1.}{\includegraphics[width=9cm]{fig2bis}}
\scalebox{1.}{\hspace{12mm} \rotatebox{90}{\includegraphics[width=9cm]{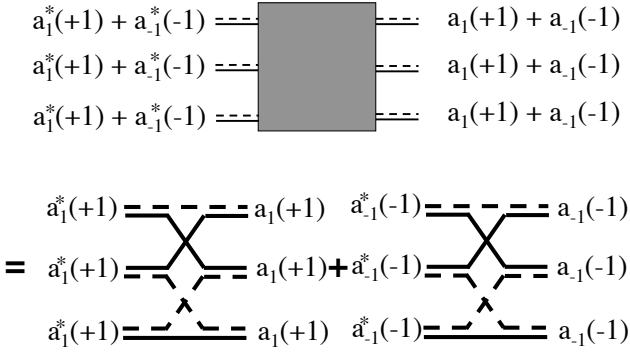}}}
\vspace{-8mm}
%\vspace{-14mm}
\caption{Shiva diagram for carrier exchanges between 3 coherent excitons
$B^\dag=a_1B_1^\dag +a_{-1}B_{-1}^\dag$. Since, according to Fig.1(a), exchanges between opposite
spin bright excitons generate dark states, this
Shiva diagram reduces to diagonal terms between $(+1)$ excitons and
between $(-1)$ excitons.}
\vspace{-6mm}
\end{figure}

This effect is enhanced in the case of polaritons. As pointed out
in \cite{22}, these mixed particles also have a photon-assisted exchange scattering, free from 
Coulomb process (see Fig.1(c)), with the same symmetry as the
Coulomb exchange scattering of Fig.1(b). This scattering is actually dominant when one of the
polaritons has a strong photon character. Consequently, the condensate of
microcavity polaritons constructed on the two bright states $(\pm 1)$, must
exhibit a linear polarization, in full agreement with experimental results
\cite{7,8}. Through a quite convincing experiment
in which excitons are created by a circularly polarized pump beam,
Snoke's group observes \cite{8} a linearly polarized light from a
region different from the excited one, thanks to a pinned geometry which
collects the excitons in a parabolic trap. In these experiments,
the exciton gas, created by a circularly polarized pump, looses its
coherence while travelling to the trap where it condenses into a
different coherent state. The linear polarization actually observed is along the
$(110)$ crystal axis, due to a small anisotropy which lifts the degeneracy
with respect to the orientation of the polarization plane, i.e. which fixes the relative phase
$\varphi_1-\varphi_{-1}$.

The situation is more complex in the case of exciton BEC because
there is no reason to eliminate the dark excitons. To characterize a Bose-Einstein condensate possibly
formed in quantum wells out of the four excitons $(\pm 1,\pm 2)$, we thus have to
look for the minimum of $\langle
H\rangle_N$ with $B^\dag$ now given by
\begin{equation}
B^\dag=a_1B_1^\dag+a_{-1}B_{-1}^\dag+a_2B_2^\dag+a_{-2}B_{-2}^\dag\ ,
\end{equation}
where $B_{\pm 2}^\dag$ reads as $B_{\pm 1}^\dag$ in Eq.\ (2), with $a_{\v
k,\mp 1/2}^\dag$ replaced by $a_{\v k,\pm 1/2}^\dag$. Since
$|a_1|^2+|a_{-1}|^2+|a_2|^2+|a_{-2}|^2=1$ for normalization, we can
split the bright and dark states as $|a_1|^2+|a_{-1}|^2=\cos^2\theta$ and
$|a_2|^2+|a_{-2}|^2=\sin^2\theta$; so that
$a_1=e^{i\varphi_1}\cos\xi_1\cos\theta$ and
$a_{-1}=e^{i\varphi_{-1}}\sin\xi_1\cos\theta$, with similar expressions
for $a_{\pm 2}$. The angles $(\theta,\xi_1,\xi_2)$ can always be taken between
$0$ and $\pi/2$. The
reduction factor in the exchange scatterings of two linear combinations of excitons, given
in Eq.\ (3), is then replaced by
\begin{eqnarray}
\hspace{-10mm}1-2|a_1a_{-1}-a_2a_{-2}|^2=1+(1+\cos\phi)\,X \nonumber \\
-\frac{1}{2} 
(\sin2\xi_1\cos^2\theta+\sin2\xi_2\sin^2\theta)^2
\end{eqnarray}
where $\phi=\varphi_2+\varphi_{-2}-\varphi_1-\varphi_{-1}$, while $X=
 \sin2\xi_1 \sin2\xi_2  \cos^2\theta\sin^2\theta$ is positive for $0\leq
(\xi_1,\xi_2)\leq\pi/2$. 
Its minimum
value with respect to $\phi$ thus occurs for $\cos\phi=-1$. The remaining factor of Eq.\ (6) is
minimum for $\xi_1=\pi/4=\xi_2$, \emph{i.e.}, when
the two bright excitons $(\pm 1)$
have the same weight in $B^\dag$, and similarly
for the two dark excitons $(\pm 2)$. However, since
for these values of $(\xi_1,\xi_2)$, the reduction factor does not depend on $\theta$,
the fraction of bright
and dark excitons in $B^\dag$, controlled by $\theta$, is not determined
by this $\langle H\rangle_N$ minimization.

We then note that, when considering the Bose-Einstein condensate
possibly formed out of the four excitons $(\pm 1,\pm 2)$, we implicitly
assume that these four excitons have the same energy. This is true if we
only take into account intraband Coulomb interactions. However
interband Coulomb processes also exist (see Fig.3(a)). While they are 
much weaker, they are crucial in the exciton BEC because they push the bright
exciton slightly above the dark exciton \cite{dark}. Indeed, these interband
processes only exist between conduction and valence electrons having the
same spin (see Fig.3(b)). Since, in quantum wells, a bright exciton $(+1)$
is made of a $(-1/2)$ electron and a $(+3/2)$ hole, i.e. a valence electron
$(-3/2)$ in an orbital state $(l_z=-1)$ and a spin state $(-1/2)$,
valence-conduction Coulomb processes do exist for bright
excitons. On the contrary, they do not exist for dark excitons:
the exciton $(S=2)$ is made of the same $(+3/2)$ hole but
its electron now has a spin $(+1/2)$. Since Coulomb interaction between electrons
is repulsive, this makes the bright exciton energy slightly above the one of the
dark exciton. Consequently, the Bose-Einstein condensate can only be formed
out of dark states. It then follows from $\langle H\rangle_N$, calculated
with $B^\dag=a_2B_2^\dag+a_{-2}B_{-2}^\dag$, that the minimum is obtained
from $|a_2|=|a_{-2}|$, i.e., excitons condense in a \emph{linearly polarized
dark state}.

If we turn to bulk samples,
the four holes $(\pm 3/2,\pm 1/2)$ now have the same energy. They give rise to
four bright excitons and four dark excitons, degenerate in
energy if we only take into account intraband Coulomb processes. However,
since the two dark excitons $(\pm 2)$ are the only ones in which
interband Coulomb interaction do not take place, these $(\pm
2)$ excitons again are the lowest energy states, out of which the condensate must
be made. Consequently, in bulk samples also, \emph{excitons condense in a linearly 
polarized dark state} made of $(\pm 2)$ excitons. 

Although the energy difference between
dark and bright excitons is quite small ($\simeq$ 0.1 meV in GaAs quantum wells), 
it is of importance to stress that it is multiplied by the number $N$ of condensed
excitons, leading to a very large total energy difference, so a
bright exciton BEC can not take place. Bright excitons however are of importance
for excited states since the dark-bright energy difference is going to be
small compared to the thermal energy.

Since dark states are not coupled to light, they cannot be seen through
luminescence. However, just for this same reason, their lifetime is much larger that
the bright exciton lifetime. This should allow to build up the density
necessary for BEC rather easily. Actually,
we are led to believe that Bose-Einstein condensates have
already been formed in some of the experiments designed to see the
exciton BEC, but they were not evidenced by lack of understanding of its dark
nature.

As a possible way to provide evidence for this dark condensate, we suggest the following.
Since dark excitons are similar to bright excitons with respect to
Pauli exclusion and Coulomb interaction, the existence of a dark
condensate must be felt by a bright exciton through a change of its
energy. This should be easily seen in a pinned geometry experiment,
similar to the one realized by Snoke's group. The spatial position of
the condensate is then known and its formation should be detected through
the change it induces to the exciton energy, as seen from the shift of the bright exciton line
emitted from this region.
\begin{figure}
\vspace{35mm}
%\vspace{-5mm}
\scalebox{1.4}{\includegraphics[width=10cm]{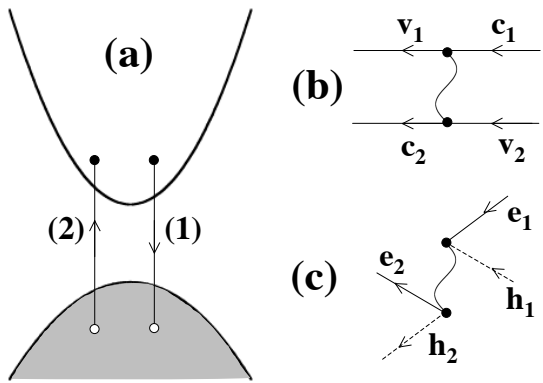}}
\vspace{-80mm}
\caption{(a) Interband valence-conduction Coulomb process. (b) Feynman
diagram for process (a); (c) same as (b) in terms of
electrons and holes: an electron-hole pair recombines while another one
is created. In these repulsive processes, the spin of the electron is
conserved; so that they do not exist for dark excitons $(\pm 2)$ made of
$(\pm1/2)$ electron and $(\pm 3/2)$ hole (\emph{i.e.}, $(\mp 3/2)$
valence electron with spin $(\mp 1/2$)).}
%\vspace{-6mm}
\end{figure}

Through a study of the time evolution of the polariton condensate polarization,
Kavokin and coworkers \cite{kavok}, have also reached the conclusion that the
condensate should
be linearly polarized. While their
work regards the dynamics of the system, we here focus
on the statics of the condensate, looking for the true ground state, through
exciton many-body effects induced by Pauli exclusion between
these composite particles. The lack of relation is also clear from the
fact that dark excitons are neglected in their works while, as we have seen, they play
a crucial role, even in the case of the polariton BEC where bright excitons are the
only ones coupled to photons.

In conclusion, our letter shows the key role played by dark excitons   
$(S=\pm 2)$ in the BEC possibly formed in semiconductors. Although they are not
generated by photon absorption, they appear in a natural way through
carrier exchanges between bright excitons with opposite spins. 
In the case of microcavity polaritons, which can
only be formed out of bright excitons, the
coherent polariton state leading to the minimum energy has a linear
polarization due to exchange couplings between bright and dark states. 
For exciton BEC, the (small)
interband valence-conduction Coulomb processes which exist for all
excitons except the ones with $(\pm 2)$ spins, make these dark excitons
slightly lower in energy. Consequently, the exciton condensate has to be
made from these $\pm2$ dark states, with again a linear polarization, 
due to the same dark-bright couplings. 
Since dark excitons have a much larger lifetime than
bright excitons, the critical density necessary for exciton Bose-Einstein
condensation should be easier to reach. However, its observation implies indirect processes
like the energy shift it induces to a bright exciton, as evidenced through the shift of the
exciton line.

We wish to thank Claude Benoit \`a la Guillaume and Roger Grousson for valuable discussions.

\end{document}